\input harvmac
\let\includefigures=\iftrue
\let\useblackboard==\iftrue
\newfam\black

\includefigures
\message{If you do not have epsf.tex (to include figures),}
\message{change the option at the top of the tex file.}
\input epsf
\def\figin{\epsfcheck\figin}\def\figins{\epsfcheck\figins}
\def\epsfcheck{\ifx\epsfbox\UnDeFiNeD
\message{(NO epsf.tex, FIGURES WILL BE IGNORED)}
\gdef\figin##1{\vskip2in}\gdef\figins##1{\hskip.5in}
\else\message{(FIGURES WILL BE INCLUDED)}%
\gdef\figin##1{##1}\gdef\figins##1{##1}\fi}
\def\DefWarn#1{}
\def\figinsert{\goodbreak\midinsert}
\def\ifig#1#2#3{\DefWarn#1\xdef#1{fig.~\the\figno}
\writedef{#1\leftbracket fig.\noexpand~\the\figno}%
\figinsert\figin{\centerline{#3}}\medskip\centerline{\vbox{
\baselineskip12pt\advance\hsize by -1truein
\noindent\footnotefont{\bf Fig.~\the\figno:} #2}}
\endinsert\global\advance\figno by1}
\else
\def\ifig#1#2#3{\xdef#1{fig.~\the\figno}
\writedef{#1\leftbracket fig.\noexpand~\the\figno}%
\global\advance\figno by1} \fi

\def\id{{1 \kern-.28em {\rm l}}}

\def\K3{{\bf K3}}
\def\journal#1&#2(#3){\unskip, \sl #1\ \bf #2 \rm(19#3) }
\def\andjournal#1&#2(#3){\sl #1~\bf #2 \rm (19#3) }

\def\bar{\overline}
\def\hat{\widehat}

\def\frac#1#2{{#1\over#2}}

\def\vev#1{\langle#1\rangle}

\def\inbar{\,\vrule height1.5ex width.4pt depth0pt}
\def\IC{\relax\hbox{$\inbar\kern-.3em{\rm C}$}}
\def\IR{\relax{\rm I\kern-.18em R}}
\def\IP{\relax{\rm I\kern-.18em P}}

%
%

%
\catcode`\@=11
\def\slash#1{\mathord{\mathpalette\c@ncel{#1}}}
\overfullrule=0pt

\def\underrel#1\over#2{\mathrel{\mathop{\kern\z@#1}\limits_{#2}}}

\catcode`\@=12


%

\def\vev#1{\left\langle #1 \right\rangle}


\lref\GoddardOD{
Goddard, P and Olive, D, eds.,
``Kac-Moody and Virasoro Algebras,"
Advanced Series in Math. Physics Vol. 3, World Scientific 1988.
}

\lref\BazhanovFT{
  V.~V.~Bazhanov, S.~L.~Lukyanov and A.~B.~Zamolodchikov,
  ``Integrable structure of conformal field theory, quantum KdV theory and thermodynamic Bethe ansatz,''
Commun.\ Math.\ Phys.\  {\bf 177}, 381 (1996).
[hep-th/9412229].
}

\lref\SmirnovLQW{
  F.~A.~Smirnov and A.~B.~Zamolodchikov,
  ``On space of integrable quantum field theories,''
Nucl.\ Phys.\ B {\bf 915}, 363 (2017).
[arXiv:1608.05499 [hep-th]].
}

\lref\LeFlochWLF{
  B.~Le Floch and M.~Mezei,
  ``KdV charges in $T\bar{T}$ theories and new models with super-Hagedorn behavior,''
SciPost Phys.\  {\bf 7}, no. 4, 043 (2019).
[arXiv:1907.02516 [hep-th]].
}

\lref\DiFrancescoNK{
  P.~Di Francesco, P.~Mathieu and D.~Senechal,
  ``Conformal Field Theory,''
  Springer-Verlag, New York, 1997.
}

\lref\MaloneyYRZ{
  A.~Maloney, G.~S.~Ng, S.~F.~Ross and I.~Tsiares,
  ``Generalized Gibbs Ensemble and the Statistics of KdV Charges in 2D CFT,''
JHEP {\bf 1903}, 075 (2019).
[arXiv:1810.11054 [hep-th]].
}

\lref\CavagliaODA{
  A.~Cavaglià, S.~Negro, I.~M.~Szécsényi and R.~Tateo,
  ``$T \bar{T}$-deformed 2D Quantum Field Theories,''
JHEP {\bf 1610}, 112 (2016).
[arXiv:1608.05534 [hep-th]].
}

\lref\CardyQAO{
  J.~Cardy,
  ``$T\bar T$ deformation of correlation functions,''
JHEP {\bf 1912}, 160 (2019), [JHEP {\bf 2019}, 160 (2020)].
[arXiv:1907.03394 [hep-th]].
}

\lref\ZamolodchikovCE{
  A.~B.~Zamolodchikov,
  ``Expectation value of composite field T anti-T in two-dimensional quantum field theory,''
[hep-th/0401146].
}

\lref\AharonyBAD{
  O.~Aharony, S.~Datta, A.~Giveon, Y.~Jiang and D.~Kutasov,
  ``Modular invariance and uniqueness of $T\bar{T}$ deformed CFT,''
JHEP {\bf 1901}, 086 (2019).
[arXiv:1808.02492 [hep-th]].
}

\lref\DymarskyETQ{
  A.~Dymarsky and K.~Pavlenko,
  ``Generalized Eigenstate Thermalization Hypothesis in 2D Conformal Field Theories,''
Phys.\ Rev.\ Lett.\  {\bf 123}, no. 11, 111602 (2019).
[arXiv:1903.03559 [hep-th]].
}

\lref\HashimotoWCT{
  A.~Hashimoto and D.~Kutasov,
  ``$T \bar{T},J \bar T$, $T \bar{J}$ Partition Sums From String Theory,''
[arXiv:1907.07221 [hep-th]].
}

\lref\SasakiMM{
  R.~Sasaki and I.~Yamanaka,
  ``Virasoro Algebra, Vertex Operators, Quantum {Sine-Gordon} and Solvable Quantum Field Theories,''
Adv.\ Stud.\ Pure Math.\  {\bf 16}, 271 (1988).
}

\lref\EguchiHS{
  T.~Eguchi and S.~K.~Yang,
  ``Deformations of Conformal Field Theories and Soliton Equations,''
Phys.\ Lett.\ B {\bf 224}, 373 (1989).
}

\lref\GuicaLIA{
  M.~Guica,
  ``An integrable Lorentz-breaking deformation of two-dimensional CFTs,''
SciPost Phys.\  {\bf 5}, no. 5, 048 (2018).
[arXiv:1710.08415 [hep-th]].
}

\lref\DattaTHY{
  S.~Datta and Y.~Jiang,
  ``$T\bar{T}$ deformed partition functions,''
JHEP {\bf 1808}, 106 (2018).
[arXiv:1806.07426 [hep-th]].
}

\lref\DymarskyLHF{
  A.~Dymarsky and K.~Pavlenko,
  ``Generalized Gibbs Ensemble of 2d CFTs at large central charge in the thermodynamic limit,''
JHEP {\bf 1901}, 098 (2019).
[arXiv:1810.11025 [hep-th]].
}

\lref\DymarskyIWX{
  A.~Dymarsky and K.~Pavlenko,
  ``Exact generalized partition function of 2D CFTs at large central charge,''
JHEP {\bf 1905}, 077 (2019).
[arXiv:1812.05108 [hep-th]].
}

\lref\DymarskyETQ{
  A.~Dymarsky and K.~Pavlenko,
  ``Generalized Eigenstate Thermalization Hypothesis in 2D Conformal Field Theories,''
Phys.\ Rev.\ Lett.\  {\bf 123}, no. 11, 111602 (2019).
[arXiv:1903.03559 [hep-th]].
}

\lref\GreenQT{
  M.~B.~Green,
  ``Interconnections between type II superstrings, M theory and N=4 supersymmetric Yang-Mills,''
Lect.\ Notes Phys.\  {\bf 525}, 22 (1999).
[hep-th/9903124].
}

\lref\CardySDV{
  J.~Cardy,
  ``The $ T\overline{T} $ deformation of quantum field theory as random geometry,''
JHEP {\bf 1810}, 186 (2018).
[arXiv:1801.06895 [hep-th]].
}


\Title{
} {\vbox{ \centerline{KdV Charges and the Generalized Torus Partition Sum}
\bigskip
\centerline{ in $T{\bar T}$ deformation}
}}

\bigskip
\centerline{\it Meseret Asrat}
\bigskip
\smallskip
\centerline{Enrico Fermi Institute and Department of Physics}
\centerline{University of Chicago}
 \centerline{5640 S. Ellis Av., Chicago, IL 60637, USA }
\smallskip

\vglue .3cm

\bigskip

\bigskip
\noindent

We consider KdV currents in a quantum field theory obtained by deforming a two dimensional conformal field theory via the irrelevant operator $T{\bar T}$. In this paper we determine their torus one--point functions modular properties. We find that the one--point functions decompose into a direct sum of two non--holomorphic modular forms. We also obtain a general differential equation that the KdV generalized torus partition sum satisfies. The differential equation provides a non--perturbative description of $T{\bar T}$ deformed theories.


\bigskip

\Date{}


\newsec{Introduction}

Conformal field theories in two dimensions can be characterized by two copies of an infinite dimensional symmetry algebra known as the Virasoro algebra ${\it Vir}$ \GoddardOD. The two copies are generated by the moments of the left--moving (holomorphic) $T$ and right--moving (antiholomorphic) $\bar T$ components of the energy momentum tensor.

The universal covering algebra ${\it UVir}$ generated by polynomials in the Virasoro generators contains an infinite dimensional abelian subalgebra \refs{\SasakiMM, \ \EguchiHS, \ \BazhanovFT}. The generators of this subalgebra are obtained by integrating conserved currents built from positive powers of the energy momentum tensor and its derivatives. These generators are known as the (quantum) KdV charges. We denote the left--moving (holomorphic) charges, following the notation used in \refs{\SmirnovLQW, \ \LeFlochWLF}, as $P_s$. They have the form
\eqn\kdvC{P_s = {1\over 2\pi}\oint dz T_{s + 1}(z), \quad T_{s + 1}(z) := :T^{s + 1\over 2}: + \cdots, \quad {\bar \partial}T = 0,
}
where $\cdots$ represents normal ordered lower polynomial in $T$ and its derivatives. The first left--moving (holomorphic) KdV charge $P_1$, for example, is the zero mode of the the left--moving (holomorphic) component  $T$ of the energy momentum tensor. The subscript $s$ indicates the charge's spin, and it always takes odd values \DiFrancescoNK. For convenience, we denote the right--moving (antiholomorphic) counterparts as $P_{-s}$. 

The mutually commuting KdV charges can be used to define a generalized partition function \refs{\MaloneyYRZ, \ \DymarskyIWX, \ \DymarskyLHF, \ \DymarskyETQ}. For a two dimensional conformal field theory the KdV generalized partition sum takes the form 
\eqn\partitionVG{Z\left(\left\{\nu_s\right\}\right) =  {\rm Tr} \left[e^{- 2\pi i \sum_{s}\nu_s P_s}\right],
}
where $\nu_s$ is the chemical potential for the charge $P_s$. In the large central charge limit the partition function is computed in \refs{ \DymarskyIWX, \ \DymarskyLHF}.

Under $T{\bar T}$ deformation \refs{\SmirnovLQW, \ \ZamolodchikovCE, \ \CavagliaODA} it is argued that the KdV charges can be adjusted to remain conserved, and mutually commuting along the deformation. That is, $T{\bar T}$ deformation preserves the infinite dimensional subalgebra \refs{\SmirnovLQW,\ \LeFlochWLF}. The (adjusted) left--moving KdV charges $P_s$ in the resulting quantum field theory take the form
\eqn\kdvAC{P_s = {1\over 2\pi}\oint \left(dz T_{s + 1} + d{\bar z}\Theta_{s - 1}\right), \quad \partial T_{s + 1} - {\bar \partial}\Theta_{s - 1} = 0.
}
Analogous expressions hold for the adjusted right--moving KdV charges $P_{-s}$.

In this paper we consider $T{\bar T}$ deformation of a conformal field theory on a torus. We derive a recurrence relation for the KdV charges one--point functions using their flow equations \LeFlochWLF\ and we determine their modular transformation properties. We also obtain a differential equation that the KdV generalized torus partition sum obeys.

The rest of the paper is organized as follows. In section 2 we obtain a recurrence relation for the KdV charges one--point functions at zero chemical potentials in a $T{\bar T}$ deformed quantum field theory on a torus. In the remainder of this section, using the recurrence relation, we study their one--point functions modular transformation properties. In section 3 we derive the flow equations for the KdV charges as a consistency check from the recurrence relation obtained in the previous section. In section 4 we derive a differential equation for the KdV generalized torus partition sum with all the chemical potentials turned on. In section 5 we discuss the main results and future research directions.

\newsec{KdV charges one--point functions}

We consider deforming a two dimensional conformal field theory with a $U(1)$ current on a torus via the irrelevant operator $T{\bar T}$ \refs{\SmirnovLQW, \ \ZamolodchikovCE, \ \CavagliaODA}. We first consider the torus partition sum with only the chemical potential $\nu_0 := \nu$ corresponding to the $U(1)$ current is turned on. We denote the charge corresponding to this conserved current by $P_0 := Q$.

Under the deformation it is argued that the charge $Q$ does not deform \refs{\CardyQAO, \ \LeFlochWLF}. 
Thus, the torus partition sum in the deformed theory is given by 
\eqn\partq{Z(\tau, \bar\tau, \nu|\lambda) = \sum_n e^{2\pi i\tau_1 RP_n - 2\pi \tau_2 R{\cal E}_n - 2\pi i \nu Q_n} = \sum_n e^{2\pi i\bar\tau R\vev{P_{-1}}_{n} -2\pi i\tau R\vev{P_{+1}}_{n} - 2\pi i \nu Q_n},
}
where $\tau = \tau_1 + i\tau_2$ is the complex modulus of the torus, $\lambda$ is the dimensionless coupling of the deformation, and the sum runs over all the mutual eigenstates $|n\rangle_\lambda$ of the Hamiltonian $H$, spatial momentum $P$ and $U(1)$ charge $Q$ on a circle of radius $R$. 

We denote respectively the energy, momentum and $U(1)$ charge of the state $|n\rangle_\lambda$ in the deformed theory as ${\cal E}_n(\lambda)$, $P_n$ and $Q_n$. The left and right--moving components of the Hamiltonian $H$ in the deformed state $|n\rangle_\lambda$ are given in terms of the KdV charges $P_{\pm 1}$ by
\eqn\ident{\vev{P_{\pm1}}_{n} = - {{\cal E}_n \pm P_n\over 2}.
}
The flow equations for the energy ${\cal E}_n(\lambda)$, momentum $P_n$, and $U(1)$ charge $Q_n$ are given by \refs{\SmirnovLQW, \ \ZamolodchikovCE, \ \CavagliaODA}
\eqn\spectr{\pi R{\cal E}_n (\lambda) = {1\over  \lambda }\left(\sqrt{1 + 2 \lambda \cdot  \pi R E_n +  \lambda^2\cdot  \left(\pi R P_n\right)^2} - 1\right), \quad \partial_\lambda P_n = 0, \quad \partial_\lambda Q_n = 0,
}
where $E_n$ is the energy of the undeformed state $|n\rangle$ in the original conformal field theory\foot{In our convention the dimensionless coupling $\lambda$ is related with the coupling $t$ in \CavagliaODA\ via $\lambda = {t\over \pi^2 R^2}$. The energies and momenta are related by a factor of $2\pi$, ${\cal E}^{\rm here}_n = 2\pi {\cal E}_n^{\rm there}, P^{\rm here}_n = 2\pi P_n^{\rm there}$.}.

The torus partition sum \partq\ can be shown to satisfy a recurrence relation among its expansion coefficients in the dimensionless coupling $\lambda$. We next obtain this relation. 

We first write
\eqn\partexp{R{\cal E}_n := \sum_p \lambda^p e^{(p)}_n, 
}
and we define the coefficients $Z_p$ as
\eqn\partDR{ Z := \sum_p \lambda^p Z_p.
}

At each order in the dimensionless coupling $\lambda$ in \partDR\ we make the following trades
\eqn\ident{\pi RE_n \to  -{1\over 2 }\partial_{\tau_2}, \quad \pi RP_n \to  -{i\over 2  }\partial_{\tau_1}.
}
By demanding a recurrence relation that involves only positive powers of $\tau_2$ and $\nu$ so that it is well--behaved, we find, after rewriting, the following recurrence relation for the expansion coefficients $Z_p$
\eqn\recur{Z_{p + 1} = {1\over p + 1}\left[\tau_2\left({\cal M}^{(p, p)}_{(\tau, \bar\tau)} + {p\over 2\tau_2}\partial_{\tau_2}\right)Z_p - {1\over 2}\left(\partial_{\tau_2}\tau_2 {\cal M}^{(p - 1, p - 1)}_{(\tau, \bar\tau)} + \partial_\tau \partial_{\bar\tau} \right)Z_{p - 1}\right],
}
where\foot{Note that $Z_{-1}$ = 0.}
\eqn\defOP{{\cal M}^{(p, p)}_{(\tau, \bar\tau)} = {\rm D}_{\tau}^{(p)}{\rm D}_{\bar\tau}^{(p)} - {p(p  - 1)\over 4\tau_2^2},
}
and the modular covariant derivatives \refs{\GreenQT, \ \DattaTHY} are 
\eqn\covd{{\rm D}_\tau^{(p)} = \partial_\tau - {ip\over 2\tau_2}, \quad {\rm D}_{\bar\tau}^{(p)} = \partial_{\bar\tau} + {ip\over 2\tau_2}.
}
Acting with ${\rm D}_{\tau}^{(p)}$ on a modular function or form of weight $(p, q)$ gives a modular function or form of weight $(p + 2, q)$. Similarly, acting with ${\rm D}_{\bar\tau}^{(q)}$ gives a modular function or form of weight $(p, q + 2)$ \refs{\GreenQT, \ \DattaTHY}. A non--holomorphic modular form $F^{(w, {\hat w})}$ of modular weight $(w, {\hat w})$ \GreenQT\ transforms as
\eqn\modtran{F^{(w, {\hat w})} \to F^{(w, {\hat w})}(c\tau + d)^w(c\bar\tau + d)^{\hat w},
}
under the modular transformation which acts as
\eqn\trans{\tau \to {a\tau + b\over c\tau + d},
}
with integers $a, b, c, d$, and $ad - bc = 1$. In the case in which there are no $U(1)$ currents a similar recurrence relation is obtained in \DattaTHY. 

To show that indeed the second--order linear homogeneous recurrence relation \recur\ is correct, we obtain from it the flow equations for the energy ${\cal E}_n$, momentum $P_n$, and $U(1)$ charge $Q_n$. We first rewrite \defOP\ using \covd\ as
\eqn\rewriteRC{{\cal M}^{(p, p)}_{(\tau, \bar\tau)} = \partial_\tau \partial_{\bar\tau} + {p\over 2\tau_2}\partial_{\tau_2}.
}
After substituting \rewriteRC\ in \recur\ and using \partDR\ we get
\eqn\itermREC{\eqalign{
\partial_\lambda Z  = \ &\sum_p\lambda^p\cdot (p + 1)Z_{p + 1} \cr
= &   \sum_p\left[\left(\tau_2\partial_\tau \partial_{\bar\tau} + {1\over 2}\cdot d_\lambda \cdot\partial_{\tau_2} \right) + {1\over 2}\cdot d_\lambda \cdot\partial_{\tau_2}\right]\lambda^pZ_p \cr
&- {1\over 2}\sum_p\lambda\left\{\partial_{\tau_2}\left[\tau_2\partial_\tau \partial_{\bar\tau} + {1\over 2}\cdot d_\lambda \cdot\partial_{\tau_2} \right] + \partial_\tau \partial_{\bar\tau}\right\}\lambda^{p}Z_{p},
}}
which we write as
\eqn\spectRU{\left[\left(1 - {\lambda\over 2} \partial_{\tau_2}\right)\cdot\left(\tau_2\partial_\tau \partial_{\bar\tau} + {1\over 2}\cdot  d_\lambda \cdot\partial_{\tau_2} -\partial_\lambda\right) - {\lambda\over 2} \partial_\tau \partial_{\bar\tau}\right]Z = 0, 
}
where $d_\lambda = \lambda \partial_\lambda$. Note that the power of $\tau_2$ is positive. We have from \partq
\eqn\partDRP{Z =  \sum_n  e^{2\pi i\bar\tau R\vev{P_{-1}}_{n} -2\pi i\tau R\vev{P_{+1}}_{n} - 2\pi i \nu Q_n}.
}
Using this in \spectRU\ we get the correct flow equations
\eqn\interSP{\partial_\lambda { e}_n +  \pi e_nd_\lambda e_n  + {\pi \over 2} (e_n^2 - p_n^2) = 0, \quad \partial_\lambda p_n = 0, \quad \partial_\lambda Q_n = 0. 
} 
where $e_n := R{\cal E}_n$ and $p_n := RP_n$. This form of the flow equation of the energy is known as Abel equation of the second kind.

It is shown in \CardySDV\ that the partition function in the case in which there are no $U(1)$ currents evolves according to a linear diffusion--type equation
\eqn\diff{\partial_t{\cal Z} = -\left(\partial_{L_1}\partial_{L_2'} - \partial_{L_2}\partial_{L_1'}\right){\cal Z}, \quad {\cal Z} := {Z\over A},
}
on a torus formed by identifying opposite edges of a parallelogram whose vertices lie at $(0, 0), (L_1, L_2), (L'_1, L'_2), (L_1 + L'_1, L_2+ L'_2)$. Where $A$ is the area of the torus. See the appendix for derivation of \diff\ using the flow equations \interSP. Thus, the derivation shows that the two equations \spectRU\ and \diff\ are equivalent.

We next consider turning on all the chemical potentials $\nu_s$ that couple with the KdV currents. We write the KdV generalized torus partition sum \partq\ as
\eqn\partgen{Z(\tau, \bar\tau, \nu_s|\lambda) = \sum_n e^{2\pi i\bar\tau R\vev{P_{-1}}_n - 2\pi i\tau R\vev{P_{+1}}_n - 2\pi i \nu Q_n -2\pi i \sum_{|s| = 3}\nu_s \vev{P_s}_n}.
}
We now generalize the recurrence relation \recur\ for the one--point functions of the KdV charges at zero chemical potentials. We have from their flow equations \LeFlochWLF 
\eqn\therLCC{ \vev{P_{|s|}}_n = \vev{c_{|s|}}_n\vev{P_1}_n^{|s|}, \quad \vev{P_{-|s|}}_n = \vev{c_{-|s|}}_n\vev{P_{-1}}_n^{|s|}.
}
where $\vev{c_s}_n$ are independent of $\lambda$. We look for a recurrence relation that involves only positive powers of the torus modulus $\tau_2$ so that it is well--behaved. We first write 
\eqn\kdvCE{ \vev{P_s}_n := \sum_p \lambda^p \vev{P_s}_n^{(p)}.
}
We expand the one--point function of $P_s$ in powers of the dimensionless coupling $\lambda$ as
\eqn\partgexp{Z' := -\left.\partial_{\nu_s} Z\right|_{\nu_t = 0} := \sum_p \lambda^p Z'_p, \quad t = 0, \pm 3, \pm 5, \cdots.
}
In what follows we assume $s$ is positive. For negative $s$ the analysis is exactly identical. Note that we set all the chemical potentials $\nu_t$ to zero. Thus, $Z'_0$ is a modular form of weight $(s + 1, 0)$ under modular transformation \refs{\MaloneyYRZ, \ \DymarskyETQ}. 

To begin with, we assume a recurrence relation of the form \recur\ since it reproduces the correct flow equation for the energy, that is, we start with
\eqn\recurCC{Z'_{p + 1} = {1\over p + 1}\left[\tau_2\left({\cal M}^{(p, p)}_{(\tau, \bar\tau)} + {p\over 2\tau_2}\partial_{\tau_2}\right)Z'_p - {1\over 2}\left(\partial_{\tau_2}\tau_2 {\cal M}^{(p - 1, p - 1)}_{(\tau, \bar\tau)} + \partial_\tau \partial_{\bar\tau} \right)Z'_{p - 1}\right],
}
where ${\cal M}^{(p, p)}_{(\tau, \bar\tau)}$ is given in \defOP.  We now consider the terms in the sequence to determine ${\cal M}^{(p, p)}_{(\tau, \bar\tau)}$ for the current case. We have from \partgexp
\eqn\partONE{Z'_1 = \sum_n 2\pi i \left(R\vev{P_{s}}_n^{(1)} +R\vev{P_{s}}^{(0)}_n\left(- 2\pi \tau_2 e_n^{(1)}\right)\right) e^{2\pi i\bar\tau R\vev{P_{-1}}^{(0)}_n - 2\pi i\tau R\vev{P_{+1}}^{(0)}_n }, 
}
and from \therLCC\ and \ident\ 
\eqn\expONE{R\vev{P_{s}}_n^{(1)} = s\cdot \vev{c_s}_n\left(\vev{P_{+1}}_n^{(0)}\right)^{s - 1}\cdot- {e_n^{(1)}\over 2} \  \Longrightarrow \ R\vev{P_{s}}_n^{(1)} \to R\vev{P_{s}}^{(0)}_n\cdot s \cdot -{i\over 2} \partial_{\bar\tau}.
}
Using this in \partONE\ we get
\eqn\reAR{Z'_1 =  \left(\tau_2\partial_\tau\partial_{\bar\tau} - {i\over 2}\cdot s\cdot \partial_{\bar\tau}\right)\sum_n2\pi i R \vev{P_{+1}}^{(0)}_n e^{2\pi i\bar\tau R\vev{P_{-1}}_n - 2\pi i\tau R\vev{P_{+1}}_n }, 
}
and since $Z'_0$ has modular weight $(s + 1, 0)$, we rewrite it as
\eqn\reARTCC{Z'_1 = \tau_2\left({\rm D}_{\tau}^{(s + 1)}{\rm D}_{\bar\tau}^{(0)}+ {i\over2\tau_2} {\rm D}_{\bar\tau}^{(0)}\right)Z'_0.
}

From \recurCC, on the other hand, we find 
\eqn\trIAl{Z'_1 = \tau_2\partial_\tau\partial_{\bar\tau}Z'_0,
}
and since $Z'_0$ is a modular function of weight $(s + 1, 0)$, we rewrite this as
\eqn\trIAlL{Z'_1 = \tau_2\partial_\tau\partial_{\bar\tau}Z'_0 = \left(\tau_2{\rm D}_{\tau}^{(s + 1)}{\rm D}_{\bar\tau}^{(0)} + {i\over2} {\rm D}_{\bar\tau}^{(0)} + {i\over 2} s\partial_{\bar\tau}\right)Z'_0.
}
This suggests that we need to adjust ${\cal M}^{(p, p)}_{(\tau, \bar\tau)}$ in \recur\ by adding \expONE\ as
\eqn\adjsUT{{\cal M}^{(p + s + 1, p)}_{(\tau, \bar\tau)} := {\cal M}^{(p, p)}_{(\tau, \bar\tau)} + s\cdot -{i\over 2\tau_2}\partial_{\bar\tau} = {\rm D}_{\tau}^{(p + s + 1)}{\rm D}_{\bar\tau}^{(p)} - {p(p + s - 1)  \over 4\tau_2^2} + {i\over2\tau_2} {\rm D}_{\bar\tau}^{(p)}.
}
With this \recurCC\ becomes
\eqn\recurCVC{Z'_{p + 1} = {1\over p + 1}\left[\tau_2\left({\cal M}^{(p + s + 1, p)}_{(\tau, \bar\tau)} + {p\over 2\tau_2}\partial_{\tau_2}\right)Z'_p - {1\over 2}\left(\partial_{\tau_2}\tau_2 {\cal M}^{(p + s, p - 1)}_{(\tau, \bar\tau)} + \partial_\tau \partial_{\bar\tau} \right)Z'_{p - 1}\right],
}
where now
\eqn\adjsTT{ {\cal M}^{(p + s + 1, p)}_{(\tau, \bar\tau)} =  {\rm D}_{\tau}^{(p + s + 1)}{\rm D}_{\bar\tau}^{(p)} - {p(p + s - 1)  \over 4\tau_2^2} + {i\over2\tau_2} {\rm D}_{\bar\tau}^{(p)}.
}

 The recurrence relation \recurCVC\ gives to all orders the correct terms for the one--point functions. We show this in the next section by reproducing from it the flow equations \LeFlochWLF\ for the KdV charges. In the remainder of this section we study the modular properties of the one--point functions using the recurrence relation \recurCVC.
 
 Note that $Z_1'$ \reARTCC\ has two terms with modular weights $(p + s + 1, p)$ and $(p + s, p + 1 )$ with $p = 1$. The decomposition appears to be generic to all orders in the coupling $\lambda$. We give two more examples.

From \recurCVC\ with \adjsTT\ we find that
\eqn\exZZZ{Z'_2 ={1\over 2} \left(\tau_2 {\cal M}_{(\tau, \bar\tau)}^{(s + 2, 1)}Z'_1 - {1\over 2}{\rm D}_{\tau}^{(s + 1)}\partial_{\bar\tau}Z'_0 - {i(s + 1)\over 4\tau_2}\partial_{\bar\tau}Z'_0\right).
}
We showed that $Z'_1$ has a term with modular wight $(s + 2, 1)$, this leads, thus, to terms in $Z'_2$ with modular weights $(s + 3, 2)$ and $(s + 2, 3)$. We note that \adjsTT\ can also be put into the form 
\eqn\adjsTVT{ {\cal M}^{(p + s + 1, p)}_{(\tau, \bar\tau)} =  {\rm D}_{\tau}^{(p + s)}{\rm D}_{\bar\tau}^{(p + 1)} - {p(p + s - 2)  \over 4\tau_2^2} - {i\over2\tau_2} {\rm D}_{\tau}^{(p + s)},
}
thus, the remaining term in $Z'_1$ which has the modular weight $(s + 1, 2)$ gives terms in $Z'_2$ with modular weights $(s + 2, 3)$ and $(s + 3, 2)$. Therefore, $Z'_2$ consists of only terms with modular wights $(p + s + 1, p)$ and $(p + s, p + 1)$ with $p = 2$. We next consider $Z'_3$. 

From \recurCVC\ we get
\eqn\rccZZ{Z'_3 ={1\over 3} \left(\tau_2 {\cal M}_{(\tau, \bar\tau)}^{(s + 3, 2)}Z'_2 - {1\over 4}\partial_{\tau_2}\partial_{\tau}\partial_{\bar\tau}Z'_0 - {1\over 2}\partial_\tau\partial_{\bar\tau}Z'_1\right).
}
Using the expression given in \reARTCC\ for $Z'_1$ which we rewrite it here as
\eqn\recZZZ{Z'_1 = \tau_2\partial_\tau\partial_{\bar\tau} Z'_0 - {i\over2}s\partial_{\bar\tau}Z'_0,
}
we rewrite the last two terms in \rccZZ\ as
\eqn\arCCCZ{{1\over 3}\left[{i(s + 2)\over 4}\partial_\tau \partial_{\bar\tau}^2 - {1\over 2} \tau_2\partial_{\tau}^2 {\rm D}^{(2)}_{\bar\tau}\partial_{\bar\tau} -{i\over 4\tau_2^2}\partial_{\bar\tau} - {1\over 2\tau_2}\partial_\tau \partial_{\bar\tau}\right]Z'_0.
}
This can be put into the natural form
\eqn\reaaCC{{1\over 3}\left[-{i\tau_2\over 2}{\rm D}^{(s + 3)}_{\tau}{\rm D}^{(s + 1)}_{\tau}{\rm D}^{(2)}_{\bar\tau}\partial_{\bar\tau} -{i(s + 2)\over 4}{\rm D}^{(s + 1)}_{\tau}{\rm D}^{(2)}_{\bar\tau}\partial_{\bar\tau}  + {s\over 4\tau_2}{\rm D}^{(s + 1)}_\tau \partial_{\bar\tau} + {is(s + 2)\over 8\tau_2^2} \partial_{\bar\tau}\right]Z'_0.
}
Thus, we note using \reaaCC\ in \rccZZ\ that $Z'_3$ involves only terms with modular weights $(p + s + 1, p)$ and $(p + s, p + 1)$ with $p = 3$. 

This leads naturally to conjecture that $Z'_{p + 1}$ consists of only terms with modular weights $(p + s + 2, p + 1)$ and $(p + s + 1, p + 2)$. Since the coupling has modular weight $(-1, -1)$\DattaTHY, this implies that the one--point function decomposes into the direct sum of a holomorphic modular form with modular weight $(s + 1, 0)$ and a non--holomorphic modular form with modular weight $(s, 1)$. In particular, for $s = 0$, we only have terms of modular weights $(1, 0)$ and $(0, 1)$. Thus, the one point function of a $U(1)$ charge is given by the direct sum of holomorphic modular form of weight $(1, 0)$ and antiholomorphic modular form of weight $(0, 1)$.


Similarly, by considering the one--point function of $P_{-|s|}$, we find that for negative $s$ we need to make the following adjustment 
\eqn\adJUT{ {\cal M}^{(p, p + |s| + 1)}_{(\tau, \bar\tau)} := {\cal M}^{(p, p)}_{(\tau, \bar\tau)} + |s|\cdot -{i\over 2\tau_2}\partial_{\tau} = {\rm D}_{\tau}^{(p)}{\rm D}_{\bar\tau}^{(p + |s|+1)} - {p(p + |s| - 1)\over 4\tau_2^2} - {i\over 2\tau_2}{\rm D}_{\tau}^{(p)} + {|s| + 1\over 4\tau_2^2},
}
which is the complex conjugate of \adjsUT. In this case $Z_1'$ has two terms with modular weights $(p, p + |s| + 1)$ and $(p + 1, p + |s|)$ with $p = 1$. Similarly, $Z'_{p + 1}$ involves only terms with modular weights $(p + 1, p + |s| + 2)$ and $(p + 2, p + |s| + 1)$.

In what follows we use the above recurrence relations to derive the flow equations \LeFlochWLF\ for the KdV charges and from the resulting flow equations a general differential equation that the KdV generalized torus partition sum satisfies.

\newsec{KdV charges spectrums}

In this section we obtain the flow equations for the KdV charges using the recurrence relation \recurCVC\ that we found in the previous section. We first consider the case in which $s$ is positive.

We first rewrite \adjsTT\ as
\eqn\adjsTVFT{ {\cal M}^{(p + s + 1, p)}_{(\tau, \bar\tau)} = \partial_\tau \partial_{\bar\tau} + {p\over 2\tau_2}\partial_{\tau_2} - {i\over 2\tau_2}s\partial_{\bar\tau}.
}
Using this \recurCVC\ becomes
\eqn\partnn{(p + 1)Z'_{p + 1} = \left(\tau_2\partial_\tau \partial_{\bar\tau} + p\partial_{\tau_2} - {is\over 2}\partial_{\bar\tau}\right)Z'_p -{1\over 2}  \left[\partial_{\tau_2}\left(\tau_2\partial_\tau \partial_{\bar\tau} + {(p-1)\over 2}\partial_{\tau_2} - {is\over 2}\partial_{\bar\tau}\right) + \partial_\tau \partial_{\bar\tau}\right]Z'_{p - 1}.
}
Now using \partnn\ in \partgexp\ we get
\eqn\rpatttt{\eqalign{
\partial_\lambda Z'  = \ &\sum_p\lambda^p\cdot (p + 1)Z'_{p + 1} \cr
= &   \sum_p\left[\left(\tau_2\partial_\tau \partial_{\bar\tau} +{1\over 2} d_\lambda\partial_{\tau_2} - {is\over 2}\partial_{\bar\tau}\right) + {1\over 2} d_\lambda\partial_{\tau_2}\right]\lambda^pZ_p' \cr
&- {1\over 2}\sum_p\lambda\left\{\left[\partial_{\tau_2}\left(\tau_2\partial_\tau \partial_{\bar\tau} + {1\over 2}d_\lambda\partial_{\tau_2} - {is\over 2}\partial_{\bar\tau}\right) + \partial_\tau \partial_{\bar\tau}\right]\right\}\lambda^{p}Z'_{p},
}}
which we rewrite as
\eqn\spectttt{\left[\left(1 - {\lambda\over 2} \partial_{\tau_2}\right)\cdot\left(\tau_2\partial_\tau \partial_{\bar\tau} + {1\over 2}\cdot  d_\lambda \cdot\partial_{\tau_2} - {is\over 2}\partial_{\bar\tau} -\partial_\lambda\right) - {\lambda\over 2} \partial_\tau \partial_{\bar\tau}\right]Z' = 0. 
}
We recall \partgexp\ which is
\eqn\partddt{Z' =  2\pi i\sum_n \vev{P_s}_n e^{2\pi i\bar\tau R\vev{P_{-1}}_n -2\pi i\tau R\vev{P_{+1}}_n },
}
using \partddt\ in \spectttt\ we find
\eqn\spectttrtr{\partial_\lambda \vev{P_s}_n + \pi e_n d_\lambda \vev{P_s}_n + {\pi  s \vev{P_s}_n\over 2}(e_n - p_n) = 0, \quad \partial_\lambda e_n +  \pi e_nd_\lambda e_n  + {\pi \over 2} (e_n^2 - p_n^2)= 0, \quad \partial_\lambda p_n = 0,
}
where $d_\lambda = \lambda \partial_\lambda$, $e_n = E_n R$, and $p_n = P_n R$.

For negative $s$ we get from \partgexp\ with \adJUT\ 
\eqn\partnnp{Z'_{p + 1} = \left(\tau_2\partial_\tau \partial_{\bar\tau} + p\partial_{\tau_2} - {i|s|\over 2}\partial_{\tau}\right)Z'_p -{1\over 2}  \left[\partial_{\tau_2}\left(\tau_2\partial_\tau \partial_{\bar\tau} + {(p-1)\over 2}\partial_{\tau_2} - {i|s|\over 2}\partial_{\tau}\right) + \partial_\tau \partial_{\bar\tau}\right]Z'_{p - 1},
}
from this it follows that
\eqn\spectttrtrp{\partial_\lambda \vev{P_s}_n + \pi e_n d_\lambda \vev{P_s}_n - {\pi  s \vev{P_s}_n\over 2}(e_n + p_n) = 0, \quad \partial_\lambda e_n +  \pi e_nd_\lambda e_n  + {\pi \over 2} (e_n^2 - p_n^2)= 0, \quad \partial_\lambda p_n = 0.
}

Equations \spectttrtr\ and \spectttrtrp\ can be put into the compact form 
\eqn\spectrcom{\partial_\lambda \vev{P_s}_n + \pi e_n d_\lambda \vev{P_s}_n + {\pi   \vev{P_s}_n\over 2}(|s|e_n - sp_n) = 0, \quad \partial_\lambda e_n +  \pi e_nd_\lambda e_n  + {\pi \over 2} (e_n^2 - p_n^2) = 0, \quad \partial_\lambda p_n = 0.
}
These equations are in agreement with the results obtained in \refs{\SmirnovLQW, \ \LeFlochWLF, \ \ZamolodchikovCE, \ \CavagliaODA}. This form of the differential equations is known as Abel equation of the second kind. In the next section we derive a differential equation that the generalized partition sum satisfies.

\newsec{The generalized torus partition sum}

In this section using the flow equations we obtain a differential equation that the partition sum satisfies with all the chemical potentials $\nu_s$ turned on.
The KdV generalized torus partition sum is given in \partgen\ with the charges running with the deformation coupling $\lambda$. We first note that
\eqn\partsif{\eqalign{
\left(-{1\over 2}d_\lambda \partial_{\tau_2} + \partial_\lambda\right)Z= \ & \sum_n\left[-2\pi\tau_2\left(\partial_\lambda e_n +  \pi e_nd_\lambda e_n \right) - 2\pi i \sum_s\nu_s\left(\partial_\lambda \vev{P_s}_n + \pi e_n d_\lambda \vev{P_s}_n\right)\right]\cr
& \times e^{2\pi i\bar\tau R\vev{P_{-1}}_n -2\pi i\tau R\vev{P_{+1}}_n - 2\pi i \sum_{s = 3}\nu_s \vev{P_s}_n} \cr
& + \sum_n \pi d_\lambda e_n e^{2\pi i\bar\tau R\vev{P_{-1}}_n -2\pi i\tau R\vev{P_{+1}}_n - 2\pi i \sum_{s = 3}\nu_s \vev{P_s}_n}.
}}
Making use of this we also note that the expression
\eqn\inter{-{\lambda\over 2\pi}\partial_{\tau_2}\left(-{1\over 2}d_\lambda \partial_{\tau_2} + \partial_\lambda\right)Z + {1\over \pi} \left(-{1\over 2}d_\lambda \partial_{\tau_2} + \partial_\lambda\right)Z,
}
can be written as
\eqn\parittsif{\eqalign{
& \left(-{\lambda \over 2\pi}\partial_{\tau_2} + {1\over \pi}\right) \sum_n\left[-2\pi\tau_2\left(\partial_\lambda e_n +  \pi e_nd_\lambda e_n \right) - 2\pi i \sum_s\nu_s\left(\partial_\lambda \vev{P_s}_n + \pi e_n d_\lambda \vev{P_s}_n\right)\right] \cr
& \times e^{2\pi i\bar\tau R\vev{P_{-1}}_n -2\pi i\tau R\vev{P_{+1}}_n - 2\pi i \sum_{s = 3}\nu_s \vev{P_s}_n} \cr
& + \sum_n  d_\lambda e_n \left(1 + \pi \lambda e_n\right)e^{2\pi i\bar\tau R\vev{P_{-1}}_n -2\pi i\tau R\vev{P_{+1}}_n - 2\pi i \sum_{s = 3}\nu_s \vev{P_s}_n}.
}}
Now using the flow equations for the charges in \parittsif\ we find
\eqn\pareeittsif{\eqalign{
& \left(-{\lambda \over 2\pi}\partial_{\tau_2} + {1\over \pi}\right) \sum_n\left[\pi^2\tau_2\left(e_n^2 - p_n^2\right) +  i\pi^2 \sum_s  \nu_s \vev{P_s}_n\left(|s|e_n - sp_n\right)\right] \cr
& \times e^{2\pi i\bar\tau R\vev{P_{-1}}_n -2\pi i\tau R\vev{P_{+1}}_n - 2\pi i \sum_{s = 3}\nu_s \vev{P_s}_n} \cr
& -{\pi\lambda\over 2} \sum_n   \left(e_n^2 - p_n^2\right)e^{2\pi i\bar\tau R\vev{P_{-1}}_n -2\pi i\tau R\vev{P_{+1}}_n - 2\pi i \sum_{s = 3}\nu_s \vev{P_s}_n}.
}}
It follows from \pareeittsif\ and \inter\ upon using \ident\ that the KdV generalized torus partition sum with all the chemical potentials turned on satisfies the following differential equation
\eqn\differpartt{\left({\lambda\over 2}\partial_{\tau_2} - 1\right)\left(-{1\over 2}d_\lambda \partial_{\tau_2} + \partial_\lambda - \tau_2 \partial_\tau \partial_{\bar\tau} + {i\over 2} \sum_s {1\over 2}\left(s\partial_{\tau_1} + i|s|\partial_{\tau_2}\right) d_{\nu_s}\right)Z = {\lambda \over 2}\partial_\tau \partial_{\bar\tau} Z,
}
where $d_{\nu_s} = \nu_s\partial_{\nu_s}$. Note that the powers of $\tau_2$ and $\lambda$ are positive. 

We next consider the case in which only the charge corresponding to $s = 0$ is turned on. From the above differential equation we have
\eqn\differpartt{\left({\lambda\over 2}\partial_{\tau_2} - 1\right)\left(-{1\over 2}d_\lambda \partial_{\tau_2} + \partial_\lambda - \tau_2 \partial_\tau \partial_{\bar\tau} \right)Z = {\lambda \over 2}\partial_\tau \partial_{\bar\tau} Z.
}

The $s = 0$ case is studied in the papers \refs{\DattaTHY, \ \AharonyBAD, \ \HashimotoWCT} and the partition sum is shown to satisfy the following differential equation
\eqn\dfff{\left(-{1\over 2}d_\lambda \partial_{\tau_2} + \partial_\lambda - \tau_2 \partial_\tau \partial_{\bar\tau}\right)Z = -{1\over 2\tau_2}d_\lambda Z.
}
Using equation \dfff\ in \differpartt\ we find that
\eqn\checkkk{{\lambda\over 2\tau_2}\left(-{1\over 2}d_\lambda \partial_{\tau_2} + \partial_\lambda - \tau_2 \partial_\tau \partial_{\bar\tau} + {1\over 2\tau_2}d_\lambda \right)Z = 0,
}
thus, our result is consistent with \refs{\DattaTHY, \ \AharonyBAD, \ \HashimotoWCT}.

The diffusion--type expression \diff\  now in the case in which the KdV charges chemical potentials are non--zero gets modified with reaction terms. The diffusion--reaction type equation takes the form
\eqn\diffche{\partial_t Z = -\left(\partial_{L_1}\partial_{L'_2} - \partial_{L_2}\partial_{L'_1}\right)Z + {1\over L_1^2 + L_2^2} \sum_s sF\left(L_1\partial_{L'_1} + L_2\partial_{L'_2}, \nu_s\partial_{\nu_s}\right)Z + {1\over A}G(L_1, L'_1, L_2, L'_2, t).
}
$G$ is some smooth function that is independent of the chemical potentials. It involves the partition function at zero chemical potentials. $F$ is a differential operator.

\newsec{Discussion}

In this paper we first considered the torus partition sum of a $T{\bar T}$ deformed theory with only a $U(1)$ current turned on. We obtained the recurrence relation  
\eqn\recurTY{Z_{p + 1} = {1\over p + 1}\left[\tau_2\left({\cal M}^{(p, p)}_{(\tau, \bar\tau)} + {p\over 2\tau_2}\partial_{\tau_2}\right)Z_p - {1\over 2}\left(\partial_{\tau_2}\tau_2 {\cal M}^{(p - 1, p - 1)}_{(\tau, \bar\tau)} + \partial_\tau \partial_{\bar\tau} \right)Z_{p - 1}\right],
}
where
\eqn\defOPYT{{\cal M}^{(p, p)}_{(\tau, \bar\tau)} = {\rm D}_{\tau}^{(p)}{\rm D}_{\bar\tau}^{(p)} - {p(p  - 1)\over 4\tau_2^2},
}
consistent with the flow equations of the energy, momentum and $U(1)$ charge. Note that the recurrence relation is independent of the chemical potential that couples to the current; the chemical potential does not appear explicitly. Thus, in the case in which there are no $U(1)$ currents it can be readily used to show that $Z_{p + 1}$ is a modular function of weight $(p + 1, p + 1)$ consistent with modular invariance of the partition sum \refs{\DattaTHY, \ \AharonyBAD}.

We next turned on the chemical potentials that couple to higher spin quantum KdV charges. In this case we obtained a recurrence relation for the one--point functions of the KdV charges $P_s$ at zero chemical potentials. For the left--moving charges we found that
\eqn\recurCGVC{Z'_{p + 1} = {1\over p + 1}\left[\tau_2\left({\cal M}^{(p + s + 1, p)}_{(\tau, \bar\tau)} + {p\over 2\tau_2}\partial_{\tau_2}\right)Z'_p - {1\over 2}\left(\partial_{\tau_2}\tau_2 {\cal M}^{(p + s, p - 1)}_{(\tau, \bar\tau)} + \partial_\tau \partial_{\bar\tau} \right)Z'_{p - 1}\right],
}
where
\eqn\adjsTGT{ {\cal M}^{(p + s + 1, p)}_{(\tau, \bar\tau)} =  {\rm D}_{\tau}^{(p + s + 1)}{\rm D}_{\bar\tau}^{(p)} - {p(p + s - 1)  \over 4\tau_2^2} + {i\over2\tau_2} {\rm D}_{\bar\tau}^{(p)}.
}
For the right--moving charges we showed that the corresponding recurrence relation is obtained by taking the complex conjugate of \recurCGVC.

By studying their modular properties order by order in the coupling we found that the one--point functions decompose into the direct sum of two modular forms with modular weights $(s + 1, 0)$ and $(s, 1)$ for the left--moving charges, and $(0, |s| + 1)$ and $(1, |s|)$ for the right--moving charges. We also obtained as a consistency check the flow equations  for the charges \LeFlochWLF\ using the recurrence relation \recurCGVC,
\eqn\floweq{\partial_\lambda \vev{P_s}_n + \pi e_n d_\lambda \vev{P_s}_n + {\pi   \vev{P_s}_n\over 2}(|s|e_n - sp_n) = 0,
}
where the energy ${\cal E}_n$ and momentum $P_n$ are given by
\eqn\enrmo{e_n = {\cal E}_n R = -R\left(\vev{P_{+1}}_n + \vev{P_{-1}}_n\right), \quad p_n = P_n R = -R\left(\vev{P_{+1}}_n - \vev{P_{-1}}_n\right),
}
which using \floweq\ give \refs{\SmirnovLQW, \ \ZamolodchikovCE, \ \CavagliaODA}
\eqn\energymo{\partial_\lambda e_n +  \pi e_nd_\lambda e_n  + {\pi \over 2} (e_n^2 - p_n^2) = 0, \quad \partial_\lambda p_n = 0.
}

We obtained also a general differential equation that the KdV generalized torus partition sum satisfies. The KdV generalized torus partition sum with all the chemical potentials turned on,
\eqn\parti{Z(\{ \nu_s\}|\lambda) =  \sum_n e^{- 2\pi i \sum_{s}\nu_s \vev{P_s}_n},
}
satisfies the differential equation 
\eqn\differparttd{\eqalign{
\left[{i\over 2}\lambda\left(\partial_{\nu_1} + \partial_{\nu_{-1}}\right) - 1\right] \left\{-{i\over 2}d_\lambda\left(\partial_{\nu_1} + \partial_{\nu_{-1}}\right) + \partial_\lambda - {i\over 2}(\nu_1 + \nu_{-1})\partial_{\nu_1}\partial_{\nu_{-1}} \right. \cr
\left. + {i\over 2}\sum_s d_{\nu_s}\left[\left({s - |s|\over 2}\right)\partial_{\nu_1} - \left({s + |s|\over 2}\right)\partial_{\nu_{-1}}\right]\right\}Z = -{\lambda\over 2}\partial_{\nu_1}\partial_{\nu_{-1}}Z,
}}
where $\nu_1 = \tau, \ \nu_{-1} = -\bar\tau, \ d_\lambda = \lambda \partial_\lambda, \ d_{\nu_s} = \nu_s\partial_{\nu_s}$. Here $s$ takes the values $0, \pm 1, \pm 3, \cdots$. The chemical potential $\nu_0$ couples to a $U(1)$ charge. The particular appearances of $\nu_1$ and $\nu_{-1}$ is due to the fact that we are deforming the conformal field theory with product of the KdV currents corresponding to $\nu_1$ and $\nu_{-1}$. The differential equation \differparttd\ can be thought as a non--perturbative description of the $T{\bar T}$ deformed theory.

In this work, we considered the KdV charges in the case in which only the $T{\bar T}$ coupling turned on. However, in general, one may consider turning on $J{\bar T}$ and $T{\bar J}$ couplings \refs{\HashimotoWCT, \ \GuicaLIA}, and it would be nice to study the modular properties  in this case.  It would be also nice to derive the flow equations for the KdV charges and also obtain the KdV generalized torus partition sum from holography. We leave these for future work. 

\bigskip\bigskip
\noindent{\bf Acknowledgements:}

I thank D. Kutasov for discussion. The work is supported in part by DOE grant DE-SC0009924.

\appendix{A}{Derivation of the diffusion equation}

In this appendix we drive the diffusion--type equation \diff\ obtained in \CardySDV\ that the torus partition function obeys using the flow equations \interSP\ which we write here for convenience,
\eqn\diffflow{
 \partial_\lambda e_n +  \pi e_nd_\lambda e_n  + {\pi \over 2} (e_n^2 - p_n^2)= 0, \quad \partial_\lambda p_n = 0.
 }
We consider a torus formed by identifying the edges of a parallelogram with vertices at $(0, 0), (L_1, L_2), (L'_1, L'_2), (L_1 + L_1', L_2 + L_2')$. The complex modulus $\tau$ of the torus is 
\eqn\difftau{\tau = {L'\over L} = \tau_1 + i\tau_2, \quad L = L_1 + iL_2, \quad L' = L'_1 + iL'_2.
}
That is,
\eqn\difftaup{\tau_1 = {1\over 2}\left({L'\over L} + {\bar L'\over \bar L}\right), \quad \tau_2 = {1\over 2i}\left({L'\over L} - {\bar L'\over \bar L}\right) = {A\over (2\pi R)^2}, \quad 
}
where $R$ and the area of the torus $A$ are given by
\eqn\diffiden{A = {i\over 2}\left(L{\bar L}' - L'{\bar L}\right), \quad (2\pi R)^2 =  L{\bar L}.
}
We recall the torus partition sum
\eqn\partqdif{Z(\tau, \bar\tau, \lambda) = \sum_n e^{2\pi i\tau_1 p_n - 2\pi \tau_2 e_n}, \quad e_n := e_n(\lambda), \quad \lambda = {t\over \pi^2 R^2}.
}
We note that the action of $\partial_\tau (\partial_{\bar\tau})$ on the torus partition sum can be identified with
\eqn\diffoper{ \partial_\tau Z(\tau, \bar\tau, \lambda) = L\partial_{L'}Z(L, L', t).
}
We also note that
\eqn\diffident{ \left(2d_t + d_R\right)e_n = 0, 
}
where $d_t = t\partial_t, d_R = R\partial_R$. Using \diffoper\ and \diffident\ we identify the action of the operator $d_t$ on the torus partition sum by
\eqn\two{d_\lambda Z = d_t Z = -{1\over 2}\left(d_L + d_{\bar L} + d_{L'} + d_{\bar L'}\right)Z,
}
where $d_L = L\partial_L, d_{\bar L} = {\bar L}\partial_{\bar L}, d_{L'} = L'\partial_{L'}, d_{\bar L'} = {\bar L'}\partial_{\bar L'}$.

Using the above relations we can now show that 

\eqn\diffushow{\eqalign{
 \sum_n\left[\pi e_n d_\lambda e_n + {\pi\over 2}(e_n^2 - p_n^2)\right] e^{2\pi i\tau_1 p_n - 2\pi \tau_2 e_n} \cr
= -{L{\bar L}\over 2\pi A}\cdot {L\bar L\over 4}\left[2i\left(\partial_{\bar L}\partial_{L'} - \partial_{L}\partial_{\bar L'}\right) - {1\over A}\left(d_L + d_{\bar L} + d_{L'} + d_{\bar L'}\right)\right]Z(L, L', t),
}}
and since
\eqn\sss{-{1\over 2\pi \tau_2}\partial_{\lambda} \equiv -{L{\bar L}\over 2\pi A}{L\bar L\over 4}\partial_t,
}
we note from \diffflow\ that 
\eqn\ggg{\partial_tZ = -\left[2i\left(\partial_{\bar L}\partial_{L'} - \partial_{L}\partial_{\bar L'}\right) - {1\over A}\left(d_L + d_{\bar L} + d_{L'} + d_{\bar L'}\right)\right]Z,
}
which we can write as
\eqn\hhh{\partial_t{\cal Z} = -\left(\partial_{L_1}\partial_{L_2'} - \partial_{L_2}\partial_{L_1'}\right){\cal Z}, \quad {\cal Z} := {Z\over A}.
}
In our convention the coupling $t$ is positive.

\listrefs
\end